\documentclass[a4paper, pra, twocolumn,superscriptaddress,showpacs]{revtex4}

\usepackage{amssymb}
\usepackage{amsmath}
\usepackage{epsfig}
\usepackage{color}
\usepackage{graphics, graphicx}
\usepackage{bbold}
\usepackage{psfrag}
\usepackage{mathcomp}
\usepackage{subfigure}
\usepackage{verbatim}
\usepackage{color}
\usepackage[colorlinks,citecolor=blue]{hyperref}

\begin{document}

\date{\today}
\title{Topological superradiant state in Fermi gases with cavity induced spin-orbit coupling}
\author{Dongyang Yu}
\affiliation{Department of Physics, Renmin University of China, Beijing 100872, People's Republic of China}
\author{Jian-Song Pan}
\affiliation{Key Laboratory of Quantum Information, University of Science and Technology of China, CAS, Hefei, Anhui 230026, People's Republic of China}
\affiliation{Synergetic Innovation Center of Quantum Information and Quantum Physics, University of Science and Technology of China, Hefei, Anhui 230026, China}
\author{Xiong-Jun Liu}
\email{phyliuxiongjun@gmail.com}
\affiliation{International Center for Quantum Materials, School of Physics, Peking University, Beijing 100871, China}
\affiliation{Collaborative Innovation Center of Quantum Matter, Beijing 100871, China}
\author{Wei Zhang}
\email{wzhangl@ruc.edu.cn}
\affiliation{Department of Physics, Renmin University of China, Beijing 100872, People's Republic of China}
\affiliation{Beijing Key Laboratory of Opto-electronic Functional Materials and Micro-nano Devices,
Renmin University of China, Beijing 100872, China}
\author{Wei Yi}
\email{wyiz@ustc.edu.cn}
\affiliation{Key Laboratory of Quantum Information, University of Science and Technology of China, CAS, Hefei, Anhui 230026, People's Republic of China}
\affiliation{Synergetic Innovation Center of Quantum Information and Quantum Physics, University of Science and Technology of China, Hefei, Anhui 230026, China}

\begin{abstract}
Coherently driven atomic gases inside optical cavities hold great promise for generating rich dynamics and exotic states of matter. It was shown recently that an exotic topological superradiant state exists in a two-component degenerate Fermi gas coupled to a cavity, where local order parameters coexist with global topological invariants. In this work, we characterize in detail various properties of this exotic state, focusing on the feedback interactions between the atoms and the cavity field. In particular, we demonstrate that cavity-induced interband coupling plays a crucial role in inducing the topological phase transition between the conventional and topological superradiant states. We analyze the interesting signatures in the cavity field left by the closing and reopening of the atomic bulk gap across the topological phase boundary and discuss the robustness of the topological superradiant state by investigating the steady-state phase diagram under various conditions. Furthermore, we consider the interaction effect and discuss the interplay between the pairing order in atomic ensembles and the superradiance of the cavity mode. Our work provides many valuable insights into the unique cavity--atom hybrid system under study and is helpful for future experimental exploration of the topological superradiant state.
\end{abstract}
\pacs{67.85.Lm, 03.75.Ss, 05.30.Fk}
\maketitle

\section{Introduction}

Quantum control over atom--light interaction is at the core of many recent developments that have greatly extended the horizon of cold-atom physics. A prominent example is the experimental realization of synthetic spin--orbit coupling (SOC) in ultracold atomic gases~\cite{gauge2exp,fermisocexp1,fermisocexp2}. When the atomic states are coupled with Raman lasers, the internal and external degrees of freedom of the atoms are also coupled. This effective SOC interaction modifies the single-particle dispersion spectra and can lead to novel quantum states in both the few-body~\cite{fewbodysoc1,fewbodysoc2,fewbodysoc3} and many-body settings~\cite{socreview1,socreview2,socreview3,socreview4,socreview5,socreview6,socreview7, socreview8}. Of particular interest is the possibility of preparing and probing topological states in cold atomic gases under synthetic SOC. Although solid-state materials with topologically nontrivial properties have been extensively studied in recent years~\cite{kanereview,zhangscreview,aliceareview}, ultracold atomic gases may serve as ideal platforms for quantum simulation of exotic topological matter~\cite{topocoldatom1,topocoldatom2, topocoldatom3,topocoldatom4,xiongjunprl13,xiongjunprl14}. For instance, the highly tunable parameters of cold atoms should offer unprecedented control over interatomic interactions in a topological material and thus would provide the intriguing opportunity of simulating topological states in a strongly interacting system.

Another exciting example of quantum control over atom--light interaction is the combination of ultracold atoms and cavity quantum electrodynamics in the strong-coupling regime, where the quantum effects of both the atoms and the cavity photons become dominant~\cite{cavbecexp0,cavbecexp1,cavatomrev}. In these cavity--atom hybrid systems, an outstanding feature is the interplay between the atoms and the cavity field: As the cavity field couples the internal states of atoms and affects the collective properties of the ensemble, the ensuing dynamics of the atomic gas feeds back to the cavity photons. Hence, the cavity field not only serves as a dynamical variable for the many-body system, it also provides a means of nondestructively detecting key properties of the hybrid system. A seminal experimental achievement in these systems is the recent observation of Dicke superradiance in a Bose--Einstein condensate (BEC) coupled to cavity light fields~\cite{cavbecexp1,dicketheory1,dicketheory2,dicketheory3}. In the experiment, as the system becomes superradiant (SR), the back-action of the cavity photons on the atoms leads to self-organization of the BEC. Furthermore, it was later demonstrated that the dynamical structure factor of the BEC can also be probed by monitoring the photons leaking from the cavity~\cite{esslingernatcom}. It has been shown theoretically that when the BEC is replaced by a spinless degenerate Fermi gas, the SR cavity field opens a bulk gap at the Fermi surface, whereas the back-action of the Fermi gas gives rise to enhancement of the superradiance due to the nesting effect~\cite{cavfermion1,cavfermion2,cavfermion3}.

In this work, we study a special cavity--atom hybrid system in which the cavity field participates in Raman processes generating synthetic SOC. Thus, this scenario combines the two interesting developments mentioned above. Similar schemes have attracted much research interest recently~\cite{cavgauge1,cavgauge2,wycavity,puhancavity,atomcavity1,atomcavity2}. For a two-component BEC in a cavity, it has been shown that the cavity-assisted SOC can lead to a rich phase diagram~\cite{cavgauge1}, whereas for a two-component degenerate Fermi gas, an exotic topological superradiant (TSR) state can be stabilized~\cite{wycavity}. In the TSR state, the bulk gap opened by the SR cavity photons also topologically protects the nontrivial properties of the Fermi gas. As a result, local order parameters coexist with global topological invariants in the TSR state. Here, we extend the studies in Ref.~\cite{wycavity} by characterizing in detail various properties of the TSR state and investigating the interplay between the TSR and the superfluid (SF) orders in the presence of interatomic interactions.

At the core of our discussion is the interplay of atoms and the cavity field. We first demonstrate that cavity-induced interband coupling plays a crucial role in inducing the topological phase transition between the trivial SR and TSR states. For this purpose, we compare the phase diagrams of relevant tight-binding models involving different numbers of bands and show explicitly that by taking only the lowest several bands into consideration, one could recover, qualitatively, the phase diagram from a full-band calculation. This enables us to derive a minimum tight-binding Hamiltonian for our system, which is helpful for generalizing the steady-state phase diagram of the current system to other related systems. Importantly, we notice that in general, the multiband effect is crucial in studying Raman-induced synthetic SOC in lattice systems. This conclusion is consistent with a recent study on laser-induced SOC in lattice models~\cite{xiaolinglattice}. We then investigate the back-action of the topological phase transition of the Fermi gas on the cavity photons. Across the topological phase boundary between the trivial SR state and the TSR state, the bulk gap closes and opens again. We show that at this phase boundary, the Fermi surface at half-filling lies exactly at the gap closing point. As a result, a spin-down state on the Fermi surface would be scattered by a Raman process to the spin-up state, which is also on the Fermi surface. This is essentially the nesting effect discussed in Refs.~\cite{cavfermion1,cavfermion2,cavfermion3} for a spinless Fermi gas in a cavity. We show that the nesting effect here leads to a peak structure in the variation of the cavity photon occupation at the topological boundary, which can be used as a clear signature of the topological phase transition.

We then discuss the robustness of the TSR state by investigating the steady-state phase diagram under various conditions. We find that the TSR state as well as the general appearance of the phase diagram are robust against finite-temperature effects and that the TSR state
should persist even in the absence of a background lattice potential. We also show that when the cavity contains a larger number of atoms, which is most likely the case in experiments, the TSR state is actually more stable.
Finally, we consider the interaction between fermions and study the interplay between the SF and TSR phases. By minimizing the free energy of the atomic ensembles, we conclude that the two orders do not coexist in the ground-state phase diagram with parameters in realistic experimental conditions.

The paper is organized as follows: in Sec.~\ref{sec:model}, we outline the system configurations and our derivation of the model Hamiltonian of the quasi-one-dimensional Fermi gas. Then, in Sec.~\ref{sec:noint}, we focus on the non-interacting case to discuss the mutual feedback interactions between atoms and the cavity field, as well as the robustness of the TSR state against the effects of finite temperature, the background lattice potential, and the atom number. The interatomic interaction effect is studied in Sec.~\ref{sec:int}, where zero-temperature phase diagrams are mapped out for various interaction intensities. Finally, we summarize the study in Sec.~\ref{sec:summary}.

\section{Model}
\label{sec:model}

\begin{figure}[tbp]
\centering
\includegraphics[width=8cm]{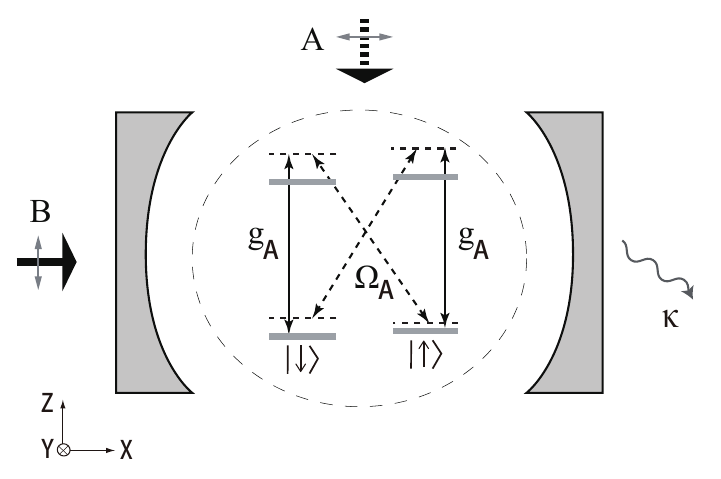}
\caption{Illustration of the system setup. A quasi-one-dimensional Fermi gas is strongly coupled to an optical cavity. The transverse pumping laser and relevant cavity mode (mode A) form two separate Raman processes coupling the two hyperfine states in the ground-state manifold of the fermions. See main text for definitions of the various parameters.}
\label{fig:config}
\end{figure}

We consider a quasi-one-dimensional, two-component Fermi gas strongly coupled to an optical cavity. While the cavity is transversely pumped by a laser field, the pumping laser and cavity field couple two hyperfine states in the ground-state manifold of each atom in two-photon Raman processes (see Fig.~\ref{fig:config}). The fermions are subject to a background lattice potential whose parameters can be controlled independently. This can be achieved, for example, by using a two-mode cavity~\cite{twomodcavexp}. While the atoms are coupled to one of the modes (pumped by A in Fig.~\ref{fig:config}), the other mode is longitudinally pumped and provides the background potential (pumped by B in Fig.~\ref{fig:config}). The effective Hamiltonian for a single atom in the Fermi gas can be written as~
\cite{wycavity}
\begin{widetext}
\begin{eqnarray}\label{equation H1}
\hat{H}_0&=&\sum_{\sigma}\int d\boldsymbol{r}\hat{\Psi}_{\sigma}^{\dagger}\left[\frac{\boldsymbol{p}^{2}}{2m}+U\left(\boldsymbol{r}\right)+
\left(V_{0}+\xi_{A}\hat{a}^{\dagger}\hat{a}\right)\cos^{2}\left(k_{0}x\right)+\xi_{\sigma}m_{z}\right]\hat{\Psi}_{\sigma} \nonumber\\
&& - \Delta_{A}\hat{a}^{\dagger}\hat{a}
+\eta\left[\int d\boldsymbol{r}\hat{\Psi}_{\uparrow}^{\dagger}\left(\hat{a}e^{ik_{0}z}+
\hat{a}^{\dagger}e^{-ik_{0}z}\right)\cos\left(k_{0}x\right)\hat{\Psi}_{\downarrow}+\rm{H.C.}\right],
\end{eqnarray}
\end{widetext}
where $\hat{\Psi}_{\sigma}$ ($\sigma=\uparrow,\downarrow$) are the fermionic field operators for different hyperfine states, $\hat{a}$ is the annihilation operator for the cavity field, $\xi_{\sigma}=\pm 1$, $m$ is the atomic mass, and $\rm{H.C.}$ represents the Hermitian conjugate. Here, $\boldsymbol{p}^{2}/2m$ is the three-dimensional kinetic energy, and $U\left(\boldsymbol{r}\right)$ is the radial trapping potential, where we consider the tight radial confinement to be in the $y$--$z$ plane. When an effective Zeeman field $m_z=\omega_{\uparrow}=-\omega_{\downarrow}$ is applied along the $z$ axis, the background lattice potential is given as $V_{0}\cos^{2}\left(k_{0}x\right)$, where $k_0$ is the wave vector of the cavity modes. The cavity detuning $\Delta_{A}=\omega_{c}-\omega_{A}$, where $\omega_c$ is the resonant frequency of cavity mode A, and $\omega_A$ is the frequency of the transverse pumping laser. We may then define $\xi_{A}=g_{A}^{2}/\Delta$, and the effective Rabi frequency of the cavity-assisted Raman processes
$\eta=\Omega_{A}g_{A}/\Delta$, where $\Delta$ is the single-photon detuning of the Raman processes.
The interaction Hamiltonian between fermions takes the form
\begin{eqnarray}
\label{eqn:Hint}
{\hat H}_{\rm int} &=&
\frac{4\pi\hbar^2 a_s}{m}
\int d {\boldsymbol r} \hat\Psi^{\dagger}_\uparrow \hat\Psi^{\dagger}_\downarrow \hat\Psi_\downarrow \hat\Psi_\uparrow,
\end{eqnarray}
where $a_s$ is the three-dimensional $s$-wave scattering length.

We assume that, under tight radial confinement, only the ground state of the radial degrees of freedom is occupied, as the Feshbach resonance is tuned within the Bardeen--Cooper--Schrieffer (BCS) side~\cite{zhang-08a, zhang-07b, foprev}. The fermionic field operator can then be written as
\begin{equation}
\hat{\Psi}_{\sigma}\left(\boldsymbol{r}\right)=\sqrt{\frac{2}{\pi\rho^{2}}}\exp\left(-\frac{y^{2}+z^{2}}{\rho^{2}}\right)\hat{\psi}_{\sigma}\left(x\right),
\end{equation}
where $\rho=\sqrt{2\hbar/m\omega_{\perp}}$ is the characteristic width of the radial harmonic confinement, with the transverse trapping frequency $\omega_{\perp}$. $\hat{\psi}_{\sigma}\left(x\right)$ is the field operator in the axial direction. If we integrate out the transverse degrees of freedom, the effective one-dimensional Hamiltonian is
\begin{eqnarray}
\label{eqn:effH}
\hat{H}_{\rm eff}&=&\sum_{\sigma}\int dx\hat{\psi}_{\sigma}^{\dagger}\bigg[\frac{p_{x}^{2}}{2m}
+ {\hat V}_{\rm eff} \cos^{2}\left(k_{0}x\right)+ \xi_{\sigma}m_{z}\bigg]\hat{\psi}_{\sigma}\nonumber\\
&&+\eta_{A}\left(\hat{a}+\hat{a}^{\dag}\right)\left[\int dx\hat{\psi}_{\uparrow}^{\dagger}\cos\left(k_{0}x\right)\hat{\psi}_{\downarrow}+\rm{H.C.}\right] \nonumber\\
&&-\Delta_A\hat{a}^{\dagger}\hat{a}
-g_{\rm 1D} \int dx \hat\psi^{\dagger}_\uparrow \hat\psi^{\dagger}_\downarrow \hat\psi_\downarrow \hat\psi_\uparrow,
\end{eqnarray}
where ${\hat V}_{\rm eff} \equiv \left(V_{0}+\xi_{A}\hat{a}^{\dag}\hat{a}\right)$, and
$\eta_{A}=s\eta$ with $s=e^{-k_{0}^{2}\rho^{2}/8}$. The effective one-dimensional interaction
strength $g_{\rm 1D}$ can be connected to the scattering length $a_s$ via the confinement-induced
resonance~\cite{olshanii-98}
\begin{equation}
\label{eqn:g1d}
g_{\rm 1D} = -\frac{4 \pi \hbar^2 a_s}{m} \frac{1}{2 \pi \rho^2} \left(1 - {\cal C} \frac{a_s}{\sqrt{2} \rho} \right)^{-1},
\end{equation}
where ${\cal C} = \lim_{s\to \infty} (\int_0^\infty ds^\prime/\sqrt{s^\prime} - \sum_{s^\prime=1}^s 1/\sqrt{s^\prime}) \approx 1.4603$. Considering the cavity decay rate $\kappa$, we can write the equation of motion for the cavity field as
\begin{eqnarray}
\label{eqn:master}
i\dot{\hat{a}}&=&\hat{a}\left[\xi_{A}\sum_{\sigma}\int dx\hat{\psi}_{\sigma}^{\dagger}\cos^{2}\left(k_{0}x\right)\hat{\psi}_{\sigma}-\Delta_{A}-i\kappa\right]\nonumber\\
&+&\eta_{A}\left[\int dx\hat{\psi}_{\downarrow}^{\dagger}\cos\left(k_{0}x\right)\hat{\psi}_{\uparrow}+\rm{H.C.}\right].
\end{eqnarray}

In the most general case, the effective Hamiltonian, Eq. (\ref{eqn:effH}), needs to be solved with the master equation for the cavity photons, Eq. (\ref{eqn:master}), to obtain the dynamical evolution of this cavity--atom hybrid system. In the limit of large photon numbers, the fluctuation of the cavity-photon operators can be neglected in a mean-field-type approach with $\hat{a} \to \alpha=\langle\hat{a}\rangle$. Furthermore, if the cavity--atom coupling is in the strong-coupling regime, the cavity field can reach a dynamical steady state within a short period of time in comparison to $1/\kappa$. Moreover, the atoms typically respond to the photon variation on an even longer time scale such that the photon dynamics can be treated adiabatically~\cite{cavbecexp0,cavbecexp1}. In this case, the steady-state solution of the system can be approximated by the stationary condition $\partial\alpha/\partial t=0$~\cite{cavfermion1,cavfermion2,cavfermion3}, which leads to
\begin{equation}
\label{eqn:alpha}
\alpha=\frac{\eta_{A}\int dx\cos\left(k_{0}x\right)\left[\langle\hat{\psi}_{\downarrow}^{\dagger}\hat{\psi}_{\uparrow}\rangle+\rm{H.C.}\right]}{\Delta_{A}+i\kappa-\xi_{A}\sum_{\sigma}\int dx\langle\hat{\psi}_{\sigma}^{\dagger}\hat{\psi}_{\sigma}\rangle\cos^{2}\left(k_{0}x\right)}.
\end{equation}
Apparently, when the system is in the steady state, the cavity mean field $\alpha$ depends on the spin--spin correlation $\langle\hat{\psi}_{\downarrow}^{\dagger}\hat{\psi}_{\uparrow}\rangle$ of the Fermi gas, whereas the latter is affected in turn by the cavity mean field and the interatomic interaction through the effective Hamiltonian, Eq.~(\ref{eqn:effH}). These observations clearly demonstrate the feedback interactions between the cavity field and the Fermi gas.

\section{Non-interacting Fermi gas and topological superradiance}
\label{sec:noint}

We first discuss the non-interacting case, where the steady state of the system can be solved self-consistently from Eqs.~(\ref{eqn:effH}) and (\ref{eqn:alpha}) with $g_{\rm 1D} = 0$. Starting from an appropriate initial cavity field $\alpha_0$, we diagonalize the effective Hamiltonian, Eq.~(\ref{eqn:effH}), for the energy spectrum. We then solve for the chemical potential from the number equation $N=\sum_{\sigma}\int dx\langle\hat{\psi}_{\sigma}^{\dagger}\hat{\psi}_{\sigma}\rangle$. The cavity mean field $\alpha$ is then updated via Eq.~(\ref{eqn:alpha}). We repeat the process until $\alpha$ converges. Note that we consider only the interesting case in which the fermions in the background lattice potential are at half-filling.

\subsection{Superradiant and topological phase transitions}\label{sec:phasediag}

As the transverse pumping strength $\eta_A$ increases past a critical value, the cavity--atom hybrid system undergoes an SR phase transition. Upon the onset of superradiance, the relevant cavity mode (mode A) becomes macroscopically occupied, and the spin--spin correlation, which serves as an order parameter, is established in the Fermi gas.

Under the adiabatic treatment of the cavity field, the SR phase transition can be characterized by the conventional Landau phase-transition theory. Starting from the Hamiltonian (\ref{eqn:effH}), we replace the cavity field operators with the mean-field value in Eq.~(\ref{eqn:alpha}). The expression for the free energy, $F=-k_BT\ln Z$, can be derived by integrating out the fermionic field operators. Here, the partition function $Z={\rm Tr}\exp[-(\hat{H}-\mu \hat{N})/k_BT]$, where $k_B$ is the Boltzmann constant, $\mu$ is the chemical potential, $\hat{N}$ is the number operator, and $T$ is the temperature.

Expanding the free energy near the critical point up to the fourth-order term in $\alpha$, we get
\begin{eqnarray}
\label{eqn:feexpand}
F&\approx&-\tilde{\Delta}_{A}\alpha^{\ast}\alpha-\chi_{\eta}\left(\alpha^{\ast}+\alpha\right)^{2}-\eta_{A}^{4}\chi_{41}\left(\alpha^{\ast}+\alpha\right)^{4}
\nonumber\\
&&-\xi^{2}\chi_{42}\left(\alpha^{\ast}\alpha\right)^{2}
+\eta_{A}^{2}\xi^{2}_{A}\chi_{43}\left(\alpha^{\ast}\alpha\right)\left(\alpha^{\ast}+\alpha\right)^{2},
\end{eqnarray}
where
\begin{widetext}
\begin{eqnarray}
\tilde{\Delta}_{A}&=&\Delta_{A}-V_{0}\sum_{j}V_{jj}n_{F}\left(\epsilon_{j}\right),\\
\chi_{\eta}&=&\eta_{A}^{2}f=-\frac{\eta_{A}^{2}}{2}\sum_{j,j^{'}}\left|M_{jj^{'}}\right|^{2} \frac{n_{F}\left(\epsilon_{j}\right)-n_{F}(\epsilon_{j^{'}})}{\epsilon_{j}-\epsilon_{j^{'}}},
\label{eqn:chieta} \\
\chi_{42}&=&\frac{1}{2\xi_{A}^{2}}\sum_{k\sigma}\sum_{m,n}\left|V_{km\sigma,kn\sigma}\right|^{2} \frac{n_{F}\left(\epsilon_{km\sigma}\right)-n_{F}\left(\epsilon_{kn\sigma}\right)}{\epsilon_{kn\sigma}-\epsilon_{km\sigma}},\nonumber \\
&& \\
\chi_{41}&=&\frac{1}{\eta_{A}^{4}}\sum_{k\sigma}\sum_{m,n,o,p}n_{F} \left(\epsilon_{mk\sigma}\right)
\frac{M_{mk\sigma,nk\bar{\sigma}}M_{nk\bar{\sigma},ok\sigma }M_{ok\sigma,pk\bar{\sigma}}M_{pk\bar{\sigma},mk\sigma}}{\left(\epsilon_{nk\bar{\sigma}} -\epsilon_{mk\sigma}\right)\left(\epsilon_{ok\sigma}-\epsilon_{mk\sigma}\right)\left(\epsilon_{pk\bar{\sigma}}-\epsilon_{mk\sigma}\right)},\\
\chi_{43}&=&\frac{1}{\eta_{A}^{2}\xi_{A}}\sum_{k\sigma}\sum_{m,n,o}V_{km\sigma,kn \sigma}M_{kn\sigma,ko\bar{\sigma}}M_{,ko\bar{\sigma},km\sigma}\nonumber\\
&&
\times\bigg[\frac{n_{F}\left(\epsilon_{km\sigma}\right)}{\left(\epsilon_{kn\sigma}-\epsilon_{km\sigma}\right) \left(\epsilon_{ko\bar{\sigma}}-\epsilon_{km\sigma}\right)}
+\frac{n_{F}\left(\epsilon_{ko\bar{\sigma}}\right)} {\left(\epsilon_{km\sigma}-\epsilon_{ko\bar{\sigma}}\right)\left(\epsilon_{kn\sigma}-\epsilon_{ko\bar{\sigma}}\right)}
+\frac{n_{F}\left(\epsilon_{kn\sigma}\right)}{\left(\epsilon_{km\sigma}-\epsilon_{kn\sigma}\right) \left(\epsilon_{ko\bar{\sigma}}-\epsilon_{kn\sigma}\right)}\bigg].
\end{eqnarray}
\end{widetext}
Here, $n_{F}\left(x\right)=1/\left[e^{\left(x-\mu\right)/k_{B}T}+1\right]$ is the Fermi distribution function. The matrix elements $V_{jj'}$ and $M_{jj'}$ are
\begin{eqnarray}
V_{jj^{'}}&=&\sum_{\sigma}\int dx
\varphi_{j\sigma}^{\ast}\left(x\right)\cos^{2}\left(k_{0}x\right)\varphi_{j^{'}\sigma}\left(x\right),\\
M_{jj'}&=&{\displaystyle\sum_{\sigma\neq\sigma'}}\int dx\varphi^{\ast}_{j\sigma}\cos(k_0x)\varphi_{j'\sigma'},
\end{eqnarray}
where the subscript $j=(nk)$ is the shorthand notation for the band index $n$ and momentum $k$, and $\varphi_{j}\left(x\right)=\{\varphi_{j\uparrow},\varphi_{j\downarrow}\}^{\rm T}$ is the eigenfunction of the Hamiltonian $\hat{H}_{0}=p_{x}^{2}/2m+V_{0}\cos^{2}\left(k_{0}x\right)+m_{z}\sigma_{z}$ with the eigenenergy $\epsilon_j$.

By defining the atomic spin--spin correlation as the order parameter
\begin{equation}
\Theta=\int dx\cos\left(k_{0}x\right)\left[\langle\hat{\psi}_{\downarrow}^{\dagger}\hat{\psi}_{\uparrow}\rangle+\rm{H.C.}\right],
\end{equation}
we can rewrite the free energy expansion as
\begin{eqnarray}
\label{eqn:GL}
F&\approx&\lambda_{2}\left(\eta_{A}\Theta\right)^{2}+\lambda_{4}\left(\eta_{A}\Theta\right)^{4},
\end{eqnarray}
where the second- and fourth-order coefficients are
\begin{eqnarray}
\lambda_2 &=& -\frac{\tilde{\Delta}_{A}}{\tilde{\Delta}_{A}^{2}+\kappa^{2}}\left[1+\eta_{A}^{2}\chi_{\eta} \frac{4\tilde{\Delta}_{c}}{\tilde{\Delta}_{A}^{2}+\kappa^{2}}\right], \nonumber\\
\lambda_4 &=& -\frac{1}{\left(\tilde{\Delta}_{A}^{2}+\kappa^{2}\right)^{2}}\Bigg[\eta_{A}^{4}\chi_{41}\frac{16\tilde{\Delta}_{A}^{4}}{\left(\tilde{\Delta}_{A}^{2} +\kappa^{2}\right)^{2}}
\nonumber \\
&&+\xi_{A}^{2}\chi_{42}-\xi\eta_{A}^{2}\chi_{43}\frac{4\tilde{\Delta}_{A}^{2}}{\tilde{\Delta}_{A}^{2} +\kappa^{2}}\Bigg].
\end{eqnarray}

According to the Ginzburg--Landau free-energy functional of Eq. (\ref{eqn:GL}), the SR phase transition occurs when $\lambda_2$ vanishes, which gives the critical condition
\begin{equation}\label{eqn:critcond}
\eta_{A}^{c}=\frac{1}{2}\sqrt{\frac{\tilde{\Delta}_{A}^{2}+\kappa^{2}}{-\tilde{\Delta}_{A}f}},
\end{equation}
where $f$ is defined in Eq.~(\ref{eqn:chieta}).
When the system first enters the SR state, the cavity field $\alpha$ is still small. Hence, the steady-state cavity mean field can be obtained by minimizing the expansion in Eq.~(\ref{eqn:feexpand}) such that
\begin{equation}\label{eqn:alplt}
\alpha_{\rm GL}=\frac{1}{\tilde{\Delta}_A+i\kappa}\sqrt{\frac{-\lambda_2}{2\lambda_4}}.
\end{equation}
We confirmed numerically that the SR boundary determined by Eq.~(\ref{eqn:critcond}) coincides with that determined from a self-consistent calculation outlined in the previous section. Furthermore, we checked that whenever $\alpha$ is small, Eq.~(\ref{eqn:alplt}) agrees well with the cavity mean field from a self-consistent calculation. Although the SR phase transition can be well characterized by the Landau theory of phase transitions, we will show below that, owing to the interplay between the cavity field and the atoms, superradiance can induce a change in the topological properties of the Fermi gas under appropriate conditions.

\begin{figure}[tbp]
\centering
\includegraphics[width=8cm]{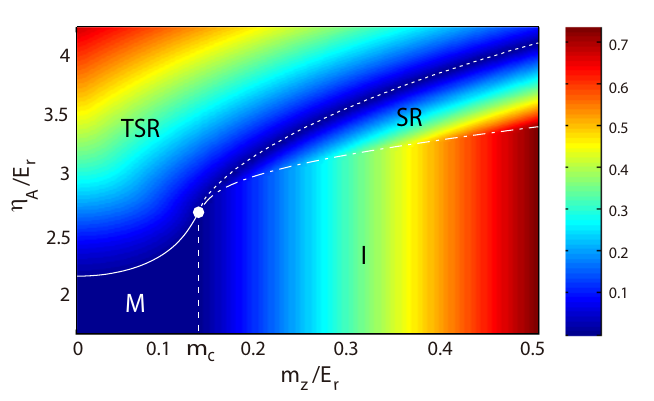}
\caption{(Color online) Steady-state phase diagram. The bulk gap of different phases is indicated by color in the background. In the absence of superradiance, the Fermi gas is in either in a metallic (M) state or an insulating (I) state, depending on the effective Zeeman field. The two phases are separated at $m_c\sim0.132E_r$. As $\eta_A$ increases, the system becomes SR, and the steady state can be either a TSR state or that of a trivial SR insulator. The TSR and SR states are further separated by a topological phase boundary at $m_z>m_c$. A tetracritical point exists at $\eta_A\sim 2.614 E_r$, $m_c\sim 0.132E_r$. We consider a half-filled lattice of $80$ sites with the parameters $k_BT=E_r/200$, $V_0=5E_r$, $\kappa= 100E_r$, $\Delta_A=-10E_r$, and $\xi_A= 5 E_r$. See the main text for the definitions of the parameters.}
\label{fig:phasediag}
\end{figure}

Whether the SR transition discussed above is topological depends on the value of the effective Zeeman field $m_z$. When $m_z$ is smaller than a critical field $m_c$, such that the Fermi gas is in a gapless metallic (M) phase before the SR transition, the system undergoes a TSR phase transition. A cavity-assisted SOC is turned on as soon as superradiance occurs. At half-filling, this SOC opens a bulk gap at the Fermi surface and mixes different spin components of the Fermi gas. The combined effects give rise to a nonzero winding number in the lowest band. The system thus features all the properties of an SR state, as well as those of a chiral topological insulator (see Fig.~\ref{fig:phasediag}). When the effective Zeeman field $m_z$ is larger than $m_c$, the Fermi gas can be in an insulating (I) state before the SR transition. Here, as the transverse pumping strength increases, an SR transition can also occur. However, as the bulk gap does not close across the SR transition, the Fermi gas remains a topologically trivial insulator. The system is then in a conventional SR state. As the pumping strength increases further, the bulk gap closes and reopens, and the system undergoes an SR--TSR phase transition (see Fig.~\ref{fig:phasediag}). In the following section, we will focus on this SR--TSR phase transition, as it is an exemplary case in which the interplay between atoms and the cavity plays a key role in shaping the properties of the system.

\subsection{Interplay between atoms and the cavity}
\label{sec:feedback}

In this subsection, we focus on the interplay between atoms and the cavity near the SR--TSR phase boundary at $m_z>m_c$. On the one hand, we show that the SR--TSR topological phase transition is induced by cavity-assisted interband coupling; on the other hand, the bulk gap closing at the topological phase boundary leaves its signature in the cavity field.

\subsubsection{Effects of cavity-assisted interband coupling}

An outstanding feature of our system is the spatial dependence of the cavity-assisted SOC, which is of odd parity with respect to the center of a lattice site. From a parity analysis, it is easy to see that the cavity-assisted SOC is capable of coupling the $s$ band to the $p$ band by a spin flip. Naturally this raises the question of the impact of higher bands. Indeed, the multiband effects play a crucial role in shaping the topological phase boundary between the SR and TSR phases.

To explicitly see the higher-band effects, we expand the effective Hamiltonian, Eq.~(\ref{eqn:effH}), onto the Wannier basis $\{\phi_{nj}\left(x\right)\}$ of the lattice Hamiltonian $p_x^2/2m+V_{\rm eff}\cos^2(k_0x)$, where $V_{\rm eff}=V_{0}+\xi_{A}\left|\alpha\right|^{2}$. Here, $n$ and $j$ are the band and site indices, respectively. The resulting tight-binding Hamiltonian can be written as
\begin{eqnarray}\label{eqn:tbH}
H_{\rm TI}&=&\sum_{n}\sum_{j,\sigma}\left[t_{s}^{n,n}\left(j,j\right)+m_{z}\xi_{\sigma}\right]\hat{\psi}_{j\sigma}^{\left(n\right)\dagger}\hat{\psi}_{j\sigma}^{\left(n\right)}
\nonumber \\
&&
+\sum_{n}\sum_{\left\langle i,j\right\rangle ,\sigma}t_{s}^{n,n}\left(i,j\right)\hat{\psi}_{i\sigma}^{\left(n\right)\dagger}\hat{\psi}_{j\sigma}^{\left(n\right)}\nonumber\\
&&+\sum_{n}\sum_{\left\langle i,j\right\rangle }\left[t_{\rm so}^{n,n}\left(i,j\right)\hat{\psi}_{i\uparrow}^{\left(n\right)\dagger}\hat{\psi}_{j\downarrow}^{\left(n\right)}+{\rm H.C.}\right]\nonumber\\
&&+\sum_{m,n }\sum_{j}\left[t_{\rm so}^{m,n}\left(j,j\right)\hat{\psi}_{j\uparrow}^{\left(m\right)\dagger}\hat{\psi}_{j\downarrow}^{\left(n\right)}+{\rm H.C.}\right],
\end{eqnarray}
where $\hat{\psi}_{j\sigma}^{\left(n\right)}$ is the atomic annihilation operator of spin $\sigma$ on site $j$ in the $n$-th band, and $\left\langle\cdot,\cdot\right\rangle$ denotes nearest-neighbor hopping. The hopping coefficients from site $i$ on the $m$-th band to site $j$ on the $n$-th band are defined as
\begin{eqnarray}
t_{s}^{m,n}(i,j)&=&\int dx\phi_{mi}^{\ast}\left(x\right)\left[\frac{p_{x}^{2}}{2m}+V_{\rm eff}\cos^{2}\left(k_{0}x\right)\right]\phi_{nj}\left(x\right),\nonumber \\
t_{\rm so}^{m,n}(i,j)&=&M_{r}\int dx\phi_{mi}^{\ast}\left(x\right)\cos\left(k_{0}x\right)\phi_{nj}\left(x\right),
\end{eqnarray}
where $M_{r}=\eta_{A}\left(\alpha^{\ast}+\alpha\right)$. Note that we have neglected the next-nearest-neighbor intraband hopping terms and the nearest-neighbor interband hopping terms, as we have checked numerically that their effects are negligible. Considering the symmetry of the Wannier functions, we have the following relations:
\begin{eqnarray}
&&t_{s}^{n,n}\left(j,j\right)=\varepsilon_{n},\quad t_{s}^{n,n}\left(j,j\pm1\right)=-\left(-1\right)^{n-1}t_{s}^{\left(n\right)},
\nonumber \\
&&t_{\rm so}^{n,n\pm1}\left(j,j\right)=\left(-1\right)^{j}t_{\rm so}^{n,n\pm1},
\nonumber \\
&&t_{\rm so}^{n,n}\left(j,j\pm1\right)=\pm\left(-1\right)^{j}t_{\rm so}^{(n)},
\end{eqnarray}
where $\varepsilon_{n}$, $t_{s}^{\left(n\right)}$, $t_{\rm so}^{n,n\pm1}$, and $t_{\rm so}^{(n)}$ are all nonnegative constants.

For a single-band tight-binding model, Eq.~(\ref{eqn:tbH}) can be simplified, as the summation over band index $n$ contains only the lowest band with $n=1$. After employing the gauge transformation $\hat{\psi}_{j\downarrow}^{\left(1\right)}\rightarrow\left(-1\right)^{j}\hat{\psi}_{j\downarrow}^{\left(1\right)}$, we can derive a concise form in momentum space~\cite{xiongjunprl13}:
\begin{equation}\label{eqn:singlebandtbm}
h^{\left(1\right)}\left(k\right)=\varepsilon_{1}+m_{z}\sigma_{z}-2t_{s}^{\left(1\right)}\cos\left(ka\right)\sigma_{z}+2t_{\rm so}^{\left(1\right)}\sigma_{y}\sin\left(ka\right).
\end{equation}
Apparently, this single-band tight-binding model is topologically nontrivial when $m_{z}<2t_{s}^{\left(1\right)}$ and is trivial otherwise. Hence, for a single-band tight-binding model, $m_c=2t_{s}^{\left(1\right)}$. Note that we have used the winding number as the topological invariant; it is defined as
\begin{equation}\label{eqn:winding}
\mathcal{W}=\sum_{\nu,\nu^{'}=y,z}\oint\frac{dk}{4\pi}\epsilon_{\nu,\nu^{'}}\hat{h}_{\nu}^{-1}\left(k\right)\partial_{k}\hat{h}_{\nu^{'}}\left(k\right).
\end{equation}
Here the integral is over the first Brillouin zone with $k\in[-k_0,k_0]$, and $\epsilon_{\nu, \nu^{'}} = \pm 1$.

\begin{figure}[tbp]
\centering
\includegraphics[width=8cm]{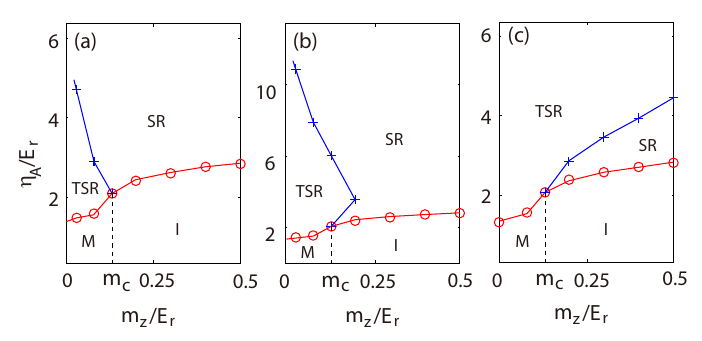}
\caption{(Color online) Phase diagram of effective tight-binding models involving (a) the lowest band, (b) the lowest two bands, and (c) the lowest three bands. Other parameters are the same as in Fig.~\ref{fig:phasediag}.}
\label{fig:multiband}
\end{figure}

In Fig.~\ref{fig:multiband}(a), we plot the steady-state phase diagram under the single-band tight-binding model, Eq.~(\ref{eqn:singlebandtbm}). Whereas the SR transition boundary (red) is determined self-consistently from Eq.~(\ref{eqn:critcond}), the SR--TSR topological phase boundary (blue) is calculated using Eq.~(\ref{eqn:winding}). Consistent with our analysis, the SR--TSR topological phase boundary is located in the region where $m_z<m_c$. Importantly, this phase boundary is qualitatively different from that under a full-band calculation, which lies at $m_z>m_c$. Thus, the coupling to higher bands can significantly alter the topological phase boundary. Although one may expect similar conclusions for Raman-assisted SOC in lattice systems in general, in our system, such higher-band effects can be seen as the back-action of atomic scattering of cavity photons.

We then characterize the steady-state phase diagram as we sequentially turn on the coupling to higher bands. More specifically, we consider the cases of two- and three-band tight-binding models. These correspond to taking $n=1,2$ and $n=1,2,3$, respectively, in the band-index summation in Eq.~(\ref{eqn:tbH}). Applying a similar gauge transformation,
$\hat{\psi}_{j\downarrow}^{\left(n\right)}\rightarrow\left(-1\right)^{j}\hat{\psi}_{j\downarrow}^{\left(n\right)}$, we can write the effective Hamiltonian in quasi-momentum space. Taking the three-band tight-binding model as an example, the Hamiltonian can be written as
\begin{equation}\label{eqn:msham}
H\left(k\right)=\left(\begin{array}{ccc}
h^{\left(1\right)}\left(k\right) & t_{\rm so}^{1,2}\sigma_{x} & 0\\
t_{\rm so}^{1,2}\sigma_{x} & h^{\left(2\right)}\left(k\right) & t_{\rm so}^{2,3}\sigma_{x}\\
0 & t_{\rm so}^{2,3}\sigma_{x} & h^{\left(3\right)}\left(k\right)
\end{array}\right),
\end{equation}
where $h^{\left(n\right)}\left(k\right)=\varepsilon_{n}+m_{z} \sigma_{z}-2\left(-1\right)^{n-1}t_{s}^{\left(n\right)}\cos\left(ka\right)\sigma_{z} +2t_{\rm so}^{\left(n\right)}\sigma_{y}\sin\left(ka\right)$.

In Figs.~\ref{fig:multiband}(b) and \ref{fig:multiband}(c), we plot the steady-state phase diagrams of the two- and three-band tight-binding models, respectively. Apparently, the SR--TSR phase boundary lies in the region of $m_z>m_c$, i.e., it is qualitatively consistent with a full-band calculation, only when the lowest three bands are considered. This suggests that a minimum tight-binding model in our system should include at least the intra- and interband coupling among the lowest three bands.

\begin{figure}[tbp]
\centering
\includegraphics[width=8cm]{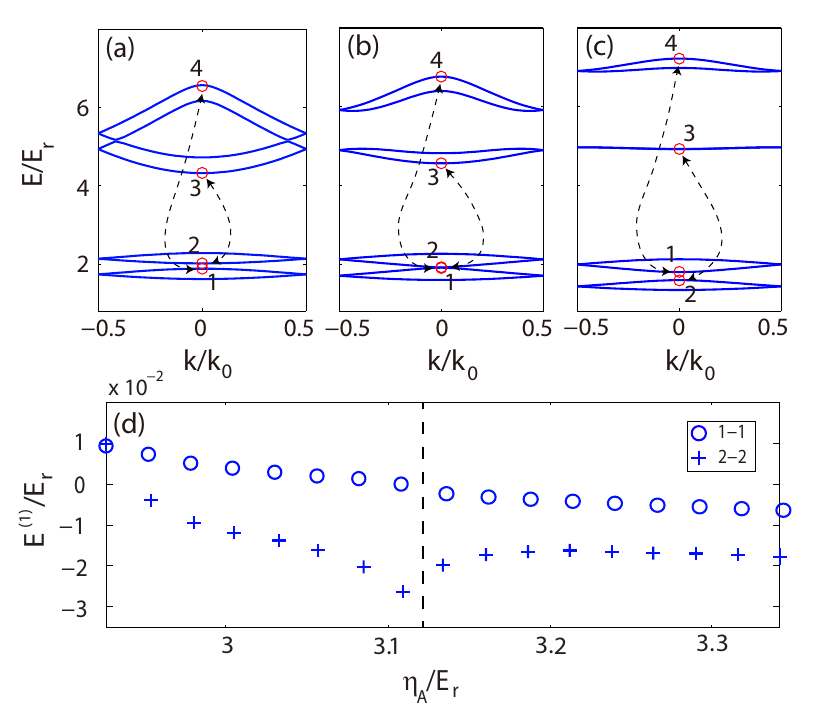}
\caption{(Color online) (a--c)Band structure and cavity-induced interband coupling for $m_z=0.2E_r>m_c$ and (a) $\eta_A\sim 2.926E_r$, (b) $\eta_A\sim 3.134E_r$, (c) $\eta_A\sim 3.342E_r$. The interband coupling among modes 1, 2, 3, and 4 at $k=0$ (red circles) is indicated by arrows. Note that in (c), owing to the band inversion at the topological boundary, the positions of labels 1 and 2 are exchanged. (d) Incremental first-order correction $E^{(1)}$, where $\delta \eta_{A,j}=0.026E_r$. 1-1 indicates the first-order correction to state 1, and 2-2 indicates the correction to state 2.}
\label{fig:bandspec}
\end{figure}

To further understand how cavity-assisted interband coupling induces the topological phase transition, we plot in Fig.~\ref{fig:bandspec} the Bloch bands in the half-folded Brillouin zone, $k\in[-k_0/2,k_0/2]$. Note that for the blue-detuned optical lattice, the zero points of the Raman coupling potential coincide with lattice sites, where the Raman potential is antisymmetric with respect to each lattice site [see Eq.~(\ref{eqn:effH})]. This symmetry leads to the following two types of interband coupling between spin-up and spin-down states at the I--SR phase boundary. First, if both the spin-up and spin-down states have the same parity (e.g., even parity for the $s,d...$ bands or odd parity for the $p,f...$ bands), the Raman potential induces only spin-flip hopping coupling but not on-site coupling. Conversely, if the spin-up and spin-down states have opposite parities, the Raman coupling potential can induce on-site spin-flip coupling as well as the hopping processes. For example,
in Fig.~\ref{fig:bandspec}(a), the states labeled 1 and 2 at the center of the Brillouin zone are coupled to the states labeled 3 and 4, respectively, via Raman processes. Importantly, this band-selective coupling persists in both the SR and TSR states, as indicated in Figs.~\ref{fig:bandspec}(b) and \ref{fig:bandspec}(c).

This interband coupling pushes each pair of Bloch bands apart. In the SR state, the band with state 2 is pushed downward faster than the band with state 1. As a result, the bulk gap between these two bands is reduced as the pumping strength $\eta_A$ increases from zero. Notice that at half-filling, the Fermi surface lies between these two bands. The bulk gap would keep decreasing until the gap closes [see Figs.~\ref{fig:bandspec}(a) and \ref{fig:bandspec}(b)]. At the gap closing point, band inversion occurs, and upon further increasing of $\eta_A$, the bulk gap is enlarged as the two bands are pushed apart by similar interband coupling mechanisms [see Fig.~\ref{fig:bandspec}(c)]. Thus, for $m_z>m_c$, the SR--TSR topological phase transition is driven by cavity-assisted interband coupling. As a result, the stability region of the TSR state is enlarged on the phase diagram.

To quantify the analysis above, we devise an incremental first-order perturbative calculation. Across the SR--TSR phase boundary, we divide the increment of the transverse pumping $\eta_A$ into $N$ small steps, which we label $\eta_{A,j}$ ($j=1,2\cdots N$). With the stepwise increase of $\eta_{A,j}$, the cavity mean-field $\alpha$ also increases, and we label it $\alpha_j$. The cavity field then affects the Hamiltonian at the $j$-th step as $H_j$, which yields the states $m_j$ [$m=1,2,3,4$, as indicated in Figs.~\ref{fig:bandspec}(a)--\ref{fig:bandspec}(c)]. At the $j$-th step, we take $m_j$ as the zeroth-order wave function and calculate the first-order correction induced by the increase in $\eta_A$: $E^{(1)}=\left\langle m_j\right|H_{j+1}-H_j\left|m_j\right\rangle$. When $\delta \eta_A=\eta_{A,j+1}-\eta_{A,j}$ is small, $E^{(1)}$ should indicate the variation in the bulk gap at $k=0$. As indicated in Fig.~\ref{fig:bandspec}(d), for $m_z<m_c$, state 2 is pushed downward, and state 1 is pushed upward. For $m_z>m_c$, the downward shift in 2 is larger than that in state 1. As discussed previously, these phenomena contribute to the bulk gap closing and reopening at the topological phase boundary. We emphasize again that the bulk gap closing and reopening are driven by cavity-assisted interband coupling, which should determine the state $m_j$ at the $j$-th step.

\subsubsection{Back-action of the cavity field across the topological phase transition}

\begin{figure}[tbp]
\centering
\includegraphics[width=8cm]{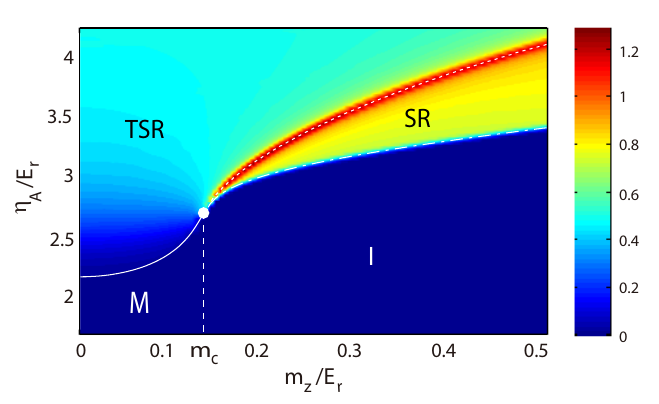}
\caption{(Color online) Steady-state phase diagram with the cavity-field variation $\Gamma$ as the background. The parameters are the same as in Fig.~\ref{fig:phasediag}.}
\label{fig:backaction}
\end{figure}

While the cavity field drives the system into the TSR state, the back-action of the topological phase transition also leaves signatures in the cavity field, which can serve as a novel nondemolition detection scheme. To demonstrate this, we define the variation of the cavity field with respect to the effective transverse pumping strength $\eta_A$: $\Gamma=\partial|\alpha|^2/\partial\eta_{A}$. In Fig.~\ref{fig:backaction}(a), we show $\Gamma$ as the background of the steady-state phase diagram. Although the value of $\Gamma$ is ostensibly higher in the SR regime than in the TSR regime, it peaks along the topological phase boundary between the two states.

Qualitatively, the peak structure in $\Gamma$ is a direct result of the bulk gap closing along the topological phase boundary. On the SR--TSR phase boundary at $m_z>m_c$, the Fermi surface at half-filling shrinks to the Dirac point at $k=0$ [see Fig.~\ref{fig:bandspec}(b)]. Therefore, a fermion close to the ``Fermi surface'' ($k=0$) interacting with photons in the cavity and the pumping fields is scattered to another state close to the Fermi surface with little energy cost. This corresponds to the nesting condition discussed in Refs.~\cite{cavfermion1,cavfermion2,cavfermion3}, under which a spinless Fermi gas coupled to a cavity field can have enhanced superradiance. Here, as the system is already in the SR state when the SR--TSR phase boundary is crossed, satisfaction of the nesting condition leads to a peak in the variation of the cavity field instead. Alternatively, the formation of the peak can be understood in terms of band inversion at the topological phase boundary. As $\eta_A$ increases past the phase boundary, the band with state 2 becomes heavily occupied owing to band inversion. As the energy difference between the band with state 2 and that with state 3 is much smaller than that between the bands with states 1 and 4, atom--photon scattering, which involves cavity-assisted interband coupling, occurs more easily when the system is in the TSR state. Thus, the abrupt change in the atom--photon scattering ability at the gap closing point gives rise to a peak in the rate of change of the average cavity photon number.

\subsection{Robustness of the topological superradiant state}\label{sec:robust}

In the previous discussions, we studied the TSR state and the relevant phase transitions in a quasi-one-dimensional Fermi gas in the presence of a background lattice potential, with a small number of atoms, and at a finite but very low temperature. An important question is the stability of the TSR state against parameters such as the temperature or background lattice potential. In this section, we discuss the effects of these parameters on the steady-state phase diagram. We also investigate the phase diagram under typical experimental conditions.

\subsubsection{Effects of temperature and background lattice potential}

For a one-dimensional Fermi gas in a cavity, the Fermi surface is just two points at zero temperature. The nesting condition is naturally satisfied when the fermions are at half-filling in the lattice potential formed by the cavity field. It has been pointed out that at zero temperature, superradiance of a one-dimensional Fermi gas can occur even with an infinitesimally small pumping field~\cite{cavfermion1,cavfermion2,cavfermion3}. As the temperature increases, the Fermi surface becomes less well-defined, which leads to a finite critical pumping field for the emergence of superradiance.

\begin{figure}[tbp]
\centering
\includegraphics[width=8cm]{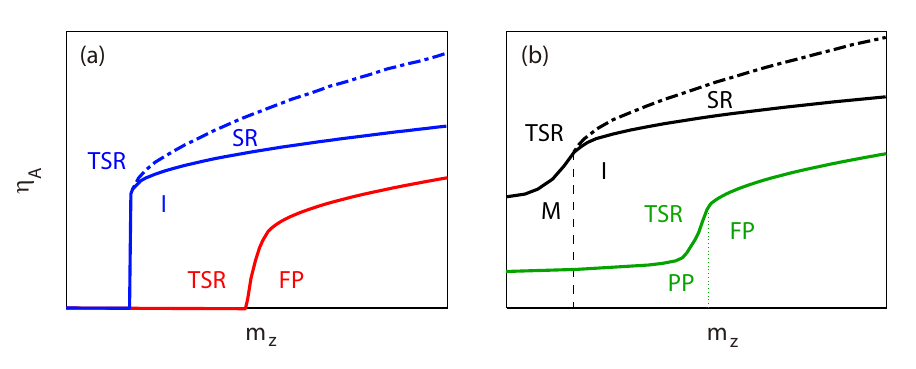}
\caption{(Color online) Illustration of typical phase diagrams with different background lattice potentials and temperatures. (a) $T=0$, with $V_0\neq0$ (blue) and $V_0=0$ (red); (b) $T\neq 0$, with $V_0\neq 0$ (black) and $V_0=0$ (green).}
\label{fig:stab}
\end{figure}

In Fig.~\ref{fig:stab}(a), we illustrate typical phase diagrams at zero temperature with (blue) and without (red) the background lattice potential $V_0$. In either case, when the effective Zeeman field $m_z$ is smaller than $m_c$, the critical pumping strength $\eta_A$ for the onset of the TSR state is indeed zero. This is consistent with previous results. When the effective Zeeman field becomes larger than $m_c$, the possible phases and phase boundaries depend on the background lattice potential. In the absence of a background lattice potential, there is no band gap before the onset of superradiance. Hence, there will be no trivial insulating state (the I or SR state), and the system enters the TSR state directly from either a fully polarized Fermi gas ($m_z>m_c$) or a partially polarized Fermi gas ($m_z<m_c$). We note that as the critical Zeeman field $m_c$ is proportional to the bandwidth, $m_c$ should increase with decreasing $V_0$. Similarly, we show in Fig.~\ref{fig:stab}(b) typical phase diagrams at finite temperatures with (black) and without (green) the background lattice potential. The main difference from the zero-temperature case is that the onset of superradiance requires a finite pumping field. Importantly, we see that the TSR state is robust against changes in the temperature and background lattice potential.

\subsubsection{Quasi-one-dimensional Fermi gases}

Quasi-one-dimensional atomic gases are typically prepared experimentally by applying a two-dimensional optical lattice potential~\cite{1dgas1}. This configuration should give rise to an array of quasi-one-dimensional atomic gases tightly confined in the transverse direction. As each quasi-one-dimensional tube in the array typically contains tens of atoms, it is worthwhile to explore the case where a large number of atoms are present in the cavity. In this case, the effective Hamiltonian should be modified as
\begin{eqnarray}\label{eqn:effHmod}
\hat{H}_{\rm eff}&=&\sum_{\nu,\sigma}\int dx\hat{\psi}_{\nu,\sigma}^{\dagger}\left[\frac{p_{x}^{2}}{2m}+
V_{\rm eff}\cos^{2}\left(k_{0}x\right)+\xi_{\sigma}m_{z}\right]\hat{\psi}_{\nu,\sigma}\nonumber\\
&&\hspace{-1cm}
+\eta_{A}\left(\alpha+\alpha^{\ast}\right)\left[\sum_{\nu}\int dx\hat{\psi}_{\nu,\uparrow}^{\dagger}\cos\left(k_{0}x\right)\hat{\psi}_{\nu,\downarrow}+\rm{H.C.}\right],
\end{eqnarray}
where $\hat{\psi}_{\nu,\sigma}$ is the atomic annihilation operator in the $\nu$-th tube. The stationary condition then gives
\begin{equation}\label{eqn:alphamod}
\alpha=\frac{\eta_{A}\sum_{\nu}\int dx\cos\left(k_{0}x\right)\left[\langle\hat{\psi}_{\nu,\downarrow}^{\dagger}\hat{\psi}_{\nu,\uparrow}\rangle+\rm{H.C.}\right]}{\Delta_{A}+i\kappa-\xi_{A}\sum_{\nu,\sigma}\int dx\langle\hat{\psi}_{\nu,\sigma}^{\dagger}\hat{\psi}_{\nu,\sigma}\rangle\cos^{2}\left(k_{0}x\right)}.
\end{equation}
It is then possible to follow the procedure outlined previously to self-consistently determine the steady state of the system and map out the phase diagram.

\begin{figure}[tbp]
\centering
\includegraphics[width=8cm]{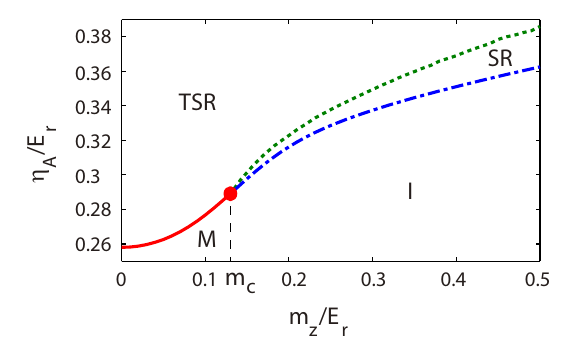}
\caption{(Color online) Phase diagram under typical experimental conditions. See discussions in the main text for details.}
\label{fig:real}
\end{figure}

As an example, we show in Fig.~\ref{fig:real} a typical phase diagram with a total number of $1.6\times 10^4$ $^6$Li atoms loaded into a two-dimensional optical lattice potential. We consider the lattice potential to have $200$ sites, which is also the number of quasi-one-dimensional tubes in the system. Compared to the previous phase diagram (see Fig.~\ref{fig:phasediag}), the tetracritical point and phase boundaries are considerably lower. More importantly, the requirements on the cavity parameters are also relaxed. For example, Fig.~\ref{fig:real} is calculated at a temperature of $k_BT=E_r/30$ ($T\sim 118$ nK for $^6$Li atoms), and with a weaker cavity--atom coupling $g_A\sim 4.3$ MHz, a larger single-photon detuning $\Delta=10$ GHz, and a larger cavity decay rate $\kappa\sim 14.8$ MHz. Hence, the TSR phase should persist under typical experimental conditions.

\section{Interacting Fermi gas and the superfluid--superradiance transition}
\label{sec:int}

In this section, we take into account the interaction effect in the Hamiltonian in Eq. (\ref{eqn:effH}) by considering an effective attractive interaction in one dimension with $g_{\rm 1D} > 0$. Notice that here we consider the strong-coupling regime where the response of cavity photons to the atomic fields can be treated adiabatically~\cite{cavbecexp0,cavbecexp1,cavfermion1,cavfermion2,cavfermion3}. By performing the gauge transformation $\hat\psi_\downarrow(x)\rightarrow-i\exp(ik_{0}x)\hat\psi_\downarrow(x)$ and adopting the mean-field formalism, we obtain the effective Hamiltonian
\begin{widetext}
\begin{eqnarray}
\label{eqn:Hmf}
\hat{H}_{\rm MF} &=&K_0 + \int dx \sum_{\sigma_1,\sigma_2} \hat\psi^\dagger_{\sigma_1}(x) \bar K_{\sigma_1,\sigma_2}(x)\hat\psi_{\sigma_2}(x) -\Delta_{A} |\alpha|^2
+ \int dx \left[ \Delta(x)\psi^{\dagger}_\uparrow(x)\psi^{\dagger}_\downarrow(x)+\Delta^{*}(x)\psi_\downarrow(x)\psi_\uparrow(x)\right],
\end{eqnarray}
where the matrix takes the form
\begin{eqnarray}
\label{eqn:K}
\bar K(x)=\Biggr[\begin{array}{cc}
\frac{p_{x}^{2}}{2m}+V(x)+m_{z} - g_{\rm 1D} n_\downarrow(x) -\mu & -iM(x)\exp(ik_{0}x) + g_{\rm 1D} R(x)\\
iM(x)\exp(-ik_{0}x) + g_{\rm 1D} R^*(x) & \frac{(p_{x}+\hbar k_{0})^{2}}{2m}+V(x)-m_{z} - g_{\rm 1D} n_\uparrow(x) - \mu
\end{array}\Biggr].
\end{eqnarray}
\end{widetext}
Here, $V(x)=(V_{0}+\xi_{A}|\alpha|^2)\cos^{2}(k_{0}x)$ is the optical lattice, $M(x)=\eta_{A}(\alpha^{*}+\alpha)\cos(k_{0}x)$ is the Raman lattice potential, and
\begin{eqnarray}
K_0=\int dx\left[g_{\rm 1D} \big(n_\uparrow(x)n_\downarrow(x)-|R(x)|^{2}\big)+\frac{|\Delta(x)|^{2}}{g_{\rm 1D}}\right]
\label{eqn:K0}
\nonumber \\
\end{eqnarray}
denotes the Hartree shift due to the interaction. In the expressions above, the density of each spin species $n_\sigma(x)=\langle \hat\psi^\dagger_\sigma(x)\hat\psi_\sigma(x) \rangle$, the SF order parameter $\Delta(x)=-g_{\rm 1D} \langle \hat\psi_\downarrow(x)\hat\psi_\uparrow(x) \rangle$, and the spin--spin correlation $R(x)=\langle \hat\psi^\dagger_\downarrow(x) \hat\psi_\uparrow(x) \rangle$.

The mean-field Hamiltonian (\ref{eqn:Hmf}) can be diagonalized by the generalized Bogoliubov transformation,
\begin{equation}
\hat\psi_\sigma(x)  =  \sum_{\eta}u_{\eta}(x\sigma)\hat C_\eta + v_{\eta}^*(x\sigma) \hat C_\eta^\dagger,
\end{equation}
leading to the Bogoliubov--de Gennes (BdG) equations,
\begin{eqnarray}
\label{eqn:bdg}
\left[
\begin{array}{cc}
\bar K(x) & \bar\Delta(x)\\
-\bar\Delta^*(x) & -\bar K^*(x)
\end{array}
\right]
\left[
\begin{array}{c}
\bar{u}_{\eta}(x)\\
\bar{v}_{\eta}(x)
\end{array}
\right]=E_{\eta}
\left[
\begin{array}{c}
\bar{u}_{\eta}(x)\\
\bar{v}_{\eta}(x)
\end{array}
\right]
\end{eqnarray}
with
\begin{eqnarray}
\bar\Delta(x) &=&
\Delta(x)\left[\begin{array}{cc}
0 & 1 \\
-1 & 0
\end{array}\right],
\nonumber \\
\bar{u}_{\eta}(x)  &=&
\left[\begin{array}{c}
u_{\eta}(x\uparrow)\\
u_{\eta}(x\downarrow)
\end{array}\right],
\bar{v}_{\eta}(x)  =
\left[\begin{array}{c}
v_{\eta}(x\uparrow)\\
v_{\eta}(x\downarrow)
\end{array}\right].
\end{eqnarray}
Here, $\hat C_\eta$ ($\hat C^\dagger_\eta$) is the annihilation (creation) operator of the quasiparticle with energy $E_\eta$, and $\bar u_\eta(x)$ ($\bar v_\eta(x)$) is the corresponding quasiparticle wave function.
Thus, the mean-field Hamiltonian for the atomic ensemble reads
\begin{eqnarray}
\hat H_{\rm MF} &=& -\sum_{\eta}E_{\eta}\int dx\sum_{\sigma}v_{\eta}^{*}(x\sigma)v_{\eta}(x\sigma)+K_{0}
\nonumber \\
&&+\sum_{\eta}E_{\eta}\hat C_\eta^\dagger\hat C_\eta.
\end{eqnarray}
Owing to particle--hole symmetry, the density for each spin component $n_\sigma(x)$, the order parameter $\Delta(x)$, and the spin--spin correlation field $R(x)$ can be expressed as
\begin{eqnarray}
n_\sigma(x) & = & \sum_{\eta} |u_{\eta}(x\sigma)|^{2} n_F (E_{\eta}),\nonumber\\
\Delta(x) & = & g_{\rm 1D} \sum_{\eta} v_{\eta}^{*}(x\uparrow)u_{\eta}(x\downarrow) n_F (E_{\eta}), \nonumber\\
R(x) &=& \sum_{\eta} u_{\eta}(x\uparrow)u_{\eta}^{*}(x\downarrow) n_F (E_{\eta}),
\end{eqnarray}
where $n_F (E)=1/(e^{E/k_BT}+1)$ is the Fermi--Dirac distribution at temperature $T$, and the summation over $\eta$ includes both the particle and hole spectra.

In the following discussion, we focus on the half-filling case, $f=1/2$, and the filling factor is defined as
\begin{equation}
f=\frac{N}{2N_0}=\frac{1}{2N_0}\int dx \sum_\sigma n_\sigma(x),
\end{equation}
where $N$ is the total particle number, and $N_0$ the number of sites. By solving the BdG equation (\ref{eqn:bdg}) for the SF order parameter $\Delta(x)$ and Eq.~(\ref{eqn:alpha}) for the cavity field $\alpha$ self-consistently, we can evaluate the energy per particle, $E_{\rm pp}=\langle \hat H_{\rm MF} \rangle/N$, for different ordered states and determine the ground state with the lowest energy. Numerically, we cut off the calculation at the lowest $N_c$ bands and employ a plane-wave approximation for high energy levels to facilitate fast convergence~\cite{liu-07}. For all the parameter regimes discussed below, convergence is obtained for $N_c \geq 13$ in a system with $N_0=80$ sites.

\begin{figure}
  \centering
  \includegraphics[width=6.1cm]{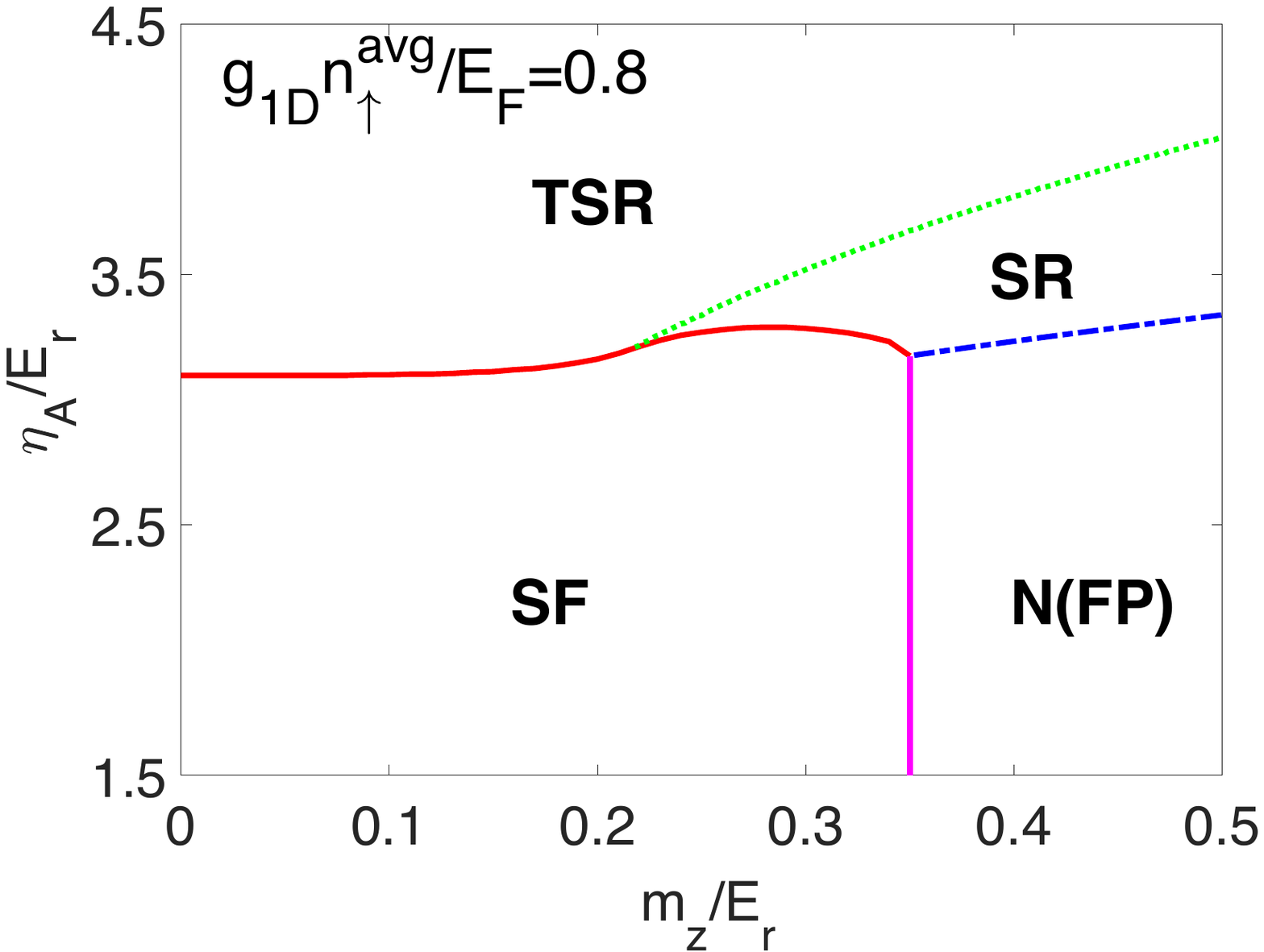}\\
  \caption{(Color online) Typical phase diagram at zero temperature with the interaction strength $g_{\rm 1D}n^{\rm avg}_\uparrow/E_F=0.8$. The SF phase is separated from the SR phases (either topologically trivial SR or TSR) and the fully polarized normal phase [N(FP)] by first-order phase transitions (solid lines). The SR--TSR and SR--N phase transitions are of the second order. Parameters used here are the same as those in Fig.~\ref{fig:phasediag}.}
  \label{fig:intphase1}
\end{figure}

\begin{figure}
  \centering
  \includegraphics[width=6.1cm]{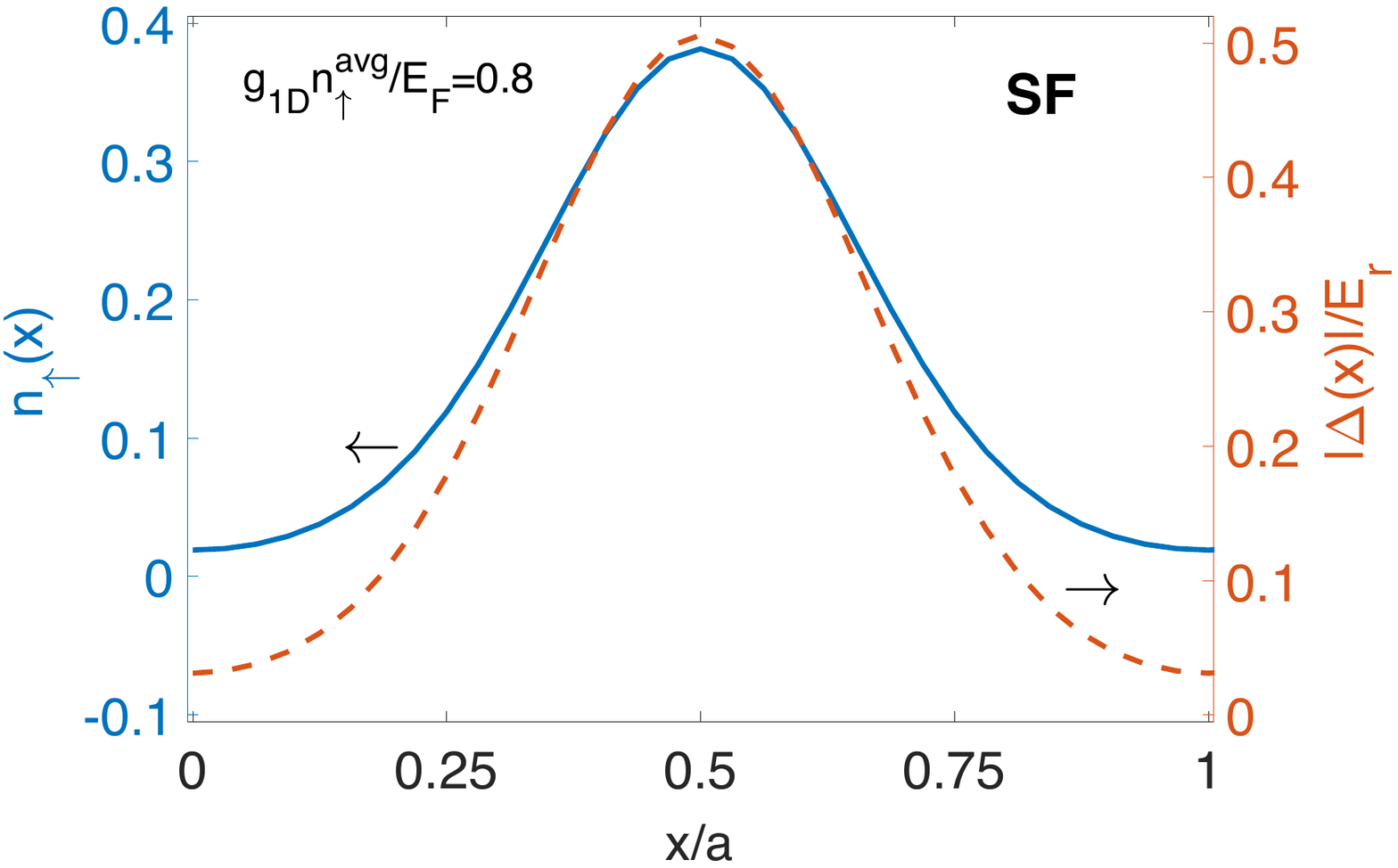}\\
  \caption{(Color online) Spatial distributions of the number density of spin-up atoms $n_{\uparrow}(x)$ and the SF order parameter $|\Delta(x)|$ within a lattice site $x \in [0,a]$. Here, $g_{\rm{1D}}n^{\rm{avg}}_\uparrow/E_F=0.8$, $m_z=0$, $\eta_A=0$, and the other parameters are the same as those in Fig.~\ref{fig:phasediag}.}
  \label{fig:density_sf}
\end{figure}

In Fig.~\ref{fig:intphase1}, we show a typical phase diagram at zero temperature for the interaction strength $g_{\rm 1D}n^{\rm avg}_\uparrow/E_F=0.8$, where $n^{\rm {avg}}_\uparrow$ is the mean density of the spin-up component, and $E_F$ denotes the Fermi energy. The attractive interaction between fermions favors an SF phase under small Zeeman fields and weak pumping powers, such that the metallic phase on the non-interacting phase diagram (Fig.~\ref{fig:phasediag}) is replaced by an SF state. The SF state can give way to the SR phase (either topologically trivial or nontrivial) as the external pumping increases. Alternatively, it gives way to a normal phase with a fully polarized Fermi surface as the Zeeman field increases. Both of these phase transitions are of the first order. The typical spatial distributions of the SF order parameter and the particle number density within a lattice site are shown in Fig.~\ref{fig:density_sf}. We emphasize that although there exist solutions for coexisting SF and SR order parameters near the first-order phase boundary between the TSR and SF states, these solutions are only metastable in energy when compared with either the SF or SR state. This result suggests that the SF and SR orders are mutually exclusive, as they couple to fermionic degrees of freedom in the particle--particle and particle--hole channels, respectively. Consequently, the topological nature of the TSR state and the topological phase transition between the TSR and SR phases are still characterized by the single-particle states of the fermions and remain the same as in Fig.~\ref{fig:phasediag}.

\begin{figure}
  \includegraphics[width=6.1cm]{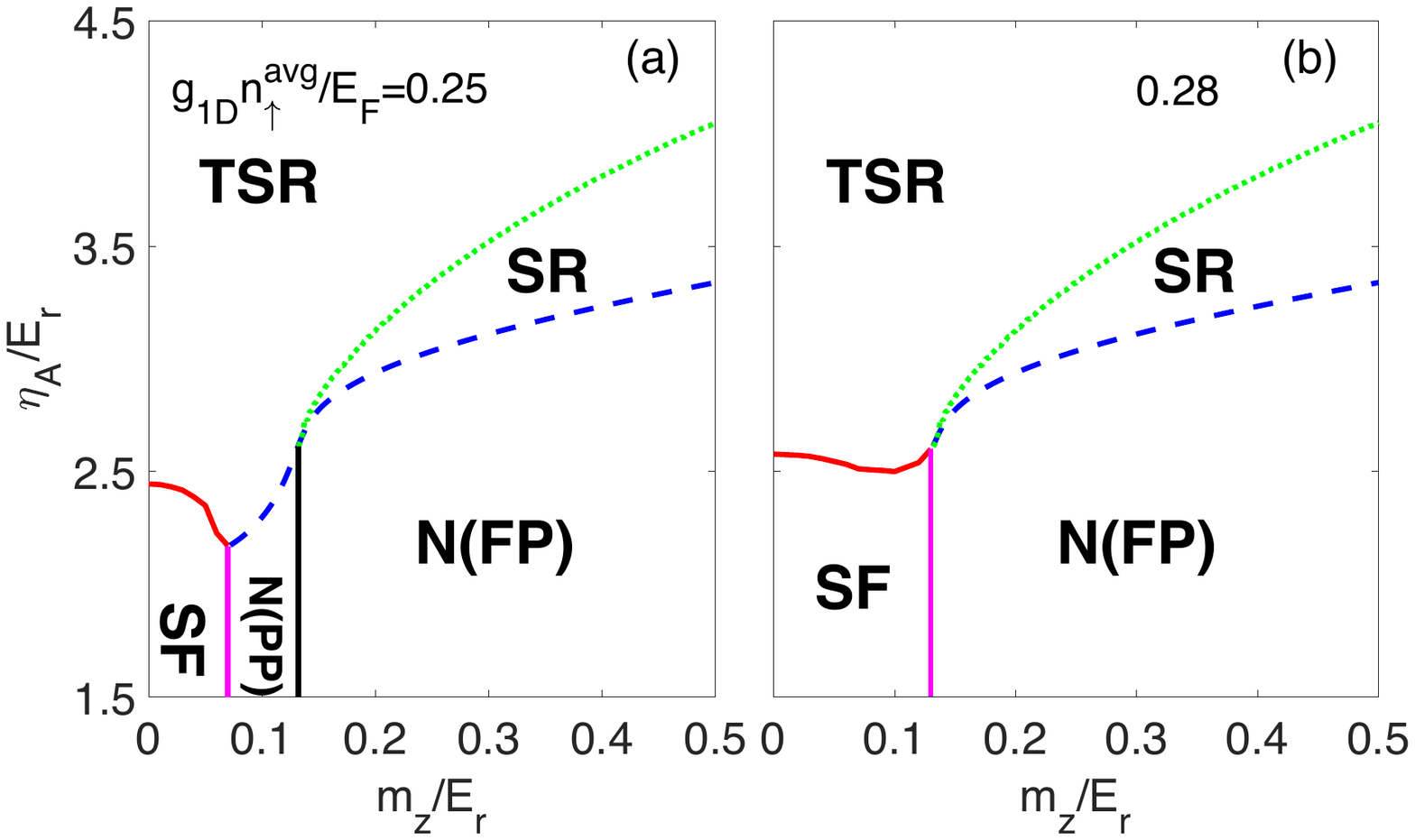}\\
  \caption{(Color online) Zero-temperature phase diagrams for (a) $g_{\rm 1D}n^{\rm avg}_\uparrow/E_F=0.25$ and (b) $g_{\rm 1D}n^{\rm avg}_\uparrow/E_F=0.28$. In the weak interaction limit, a normal phase with partially polarized (PP) Fermi surfaces is present for intermediate Zeeman field. Other parameters are the same as those in Fig.~\ref{fig:phasediag}.}
 \label{fig:intphase2}
\end{figure}

When the interaction strength becomes weaker, the SF state tends to be less favorable with a smaller condensation energy gain. Consequently, the SF--N phase transition moves toward smaller values of the Zeeman field $m_z$, extending the normal phase region. In Fig.~\ref{fig:intphase2}, we show the zero-temperature phase diagrams for $g_{\rm 1D}n^{\rm avg}_\uparrow/E_F=0.25$ and 0.28. Notice that for a weak enough interaction strength, as illustrated in Fig.~\ref{fig:intphase2}(a), the SF--N transition line occurs at a critical $m_z$ even smaller than $m_c \sim 0.132 E_r$, where the single-particle spectrum opens a spin gap. Further, because the combined effect of terms associated with the spin--spin correlation $R(x)$ in Eqs.~(\ref{eqn:K}) and (\ref{eqn:K0}) leads to a mean-field energy increase of $\sim g_{\rm 1D} |R(x)|^2$ to the energy of the SR phase, a partially polarized normal phase can be stabilized in the window between the two values of $m_z$ for a small pumping power $\eta_A$.

\section{Summary}
\label{sec:summary}

We discussed in detail various phases of a quasi-one-dimensional Fermi gas under cavity-assisted SOC. We examined an outstanding feature of the cavity--atom hybrid system, the feedback interactions between the Fermi gas and the cavity field. We explicitly demonstrated that the topological phase transition between the TSR state and the topologically trivial SR state is induced by cavity-assisted coupling to higher bands. By sequentially introducing high-lying bands into the tight-binding model, we showed that to qualitatively reproduce the topological phase boundary, a tight-binding model should include at least the lowest three bands. We also investigated the way that cavity-assisted interband coupling closes and reopens the bulk gap across the topological phase boundary. As lattice models with Raman-assisted SOC have attracted much attention recently, we expect that the effects of interband coupling on the topological phase transition are relevant to a wide class of systems. Regarding the feedback action of the topological phase transition on the cavity field, we then showed that a peak structure emerges along the topological phase boundary in the rate of change of the cavity field with respect to the pumping strength. We demonstrated that the formation of the peak structure can be understood from the perspective of either satisfaction of the nesting condition or band inversion at the bulk gap closing point. Apparently, the peak structure should be useful for nondemolition measurement of the topological phase transition. We also studied the effects of the temperature, background lattice potential, and atom number on the phase boundaries and demonstrated that the TSR state is robust under typical experimental conditions.

Finally, we studied the interplay between the SF and superradiance orders in this setup using the mean-field BdG approach. By mapping out the zero-temperature phase diagrams for various interaction strengths, we concluded that although a solution for coexisting SF and SR phases can be obtained, it is only metastable in energy in comparison to that of an SF phase without superradiance or an SR phase without pairing order. This result arises from the observation that the SF and superradiance order parameters couple to fermionic operators in the particle--particle and particle--hole channels, respectively. Consequently, the SR phase, whether topologically trivial or nontrivial, is separated from the SF phase by a first-order phase transition.

For an experimental realization, we considered a typical example of $^6$Li atoms, in which case the pseudo-spins can be chosen as $|F=1/2,m_F=\pm 1/2>$ states in the ground-state manifold. For $^{40}$K atoms, because more than two hyperfine states exist in the ground-state manifold, a large magnetic field is required to isolate the level structure shown in Fig.~\ref{fig:config}. The resulting large Zeeman splitting between the pseudo-spin states would allow the realization of only the large-$m_z$ side of the phase diagram. Another possible choice of atom species is $^{171}$Yb, whose $^1S_0$ manifold has only two states, which have nuclear spins $m_I=\pm 1/2$.

\section*{Acknowledgement}
This work is supported by the National Key Basic Research Program (2013CB922000), the National Key R\&D Program (2016YFA0301700), the National Natural Science Foundation of China (60921091, 11274009, 11374283, 11434011, 11522436, 11522545, 11574008), and the Research Funds of Renmin University of China (10XNL016, 16XNLQ03). W.Y. acknowledges support from the Strategic Priority Research Program (B) of the Chinese Academy of Sciences (XDB01030200). X.J.L. is supported by the Ministry of Science \& Technology (2016YFA0301604) and the Thousand-Young-Talent Program of China.

\end{document}